\documentclass[prl,aps,twocolumn]{revtex4}
\usepackage[]{times,amsmath,epsfig,graphicx}
\usepackage[]{graphicx,epstopdf}

\newcommand{\sech}[0]{{\,\rm sech } }

\begin{document}
	
	\title{High-contrast Kerr Frequency Combs}
	
	\author{ Ivan S. Grudinin$^{1,*}$, Vincent Huet$^1$, Nan Yu$^1$, Michael L. Gorodetsky$^{2,3}$, Andrey B. Matsko$^4$, and Lute Maleki$^4$}
	
	\affiliation{$^1$ Jet Propulsion Laboratory, California Institute of Technology, 4800 Oak Grove Drive, Pasadena, CA 91109-8099, USA}
	
	\affiliation{$^2$ Faculty of Physics, M.V. Lomonosov Moscow State University, 119991, Moscow, Russia}
	
	\affiliation{$^3$ Russian Quantum Center, Skolkovo 143025, Russia}
	
	\affiliation{$^4$  OEwaves Inc., 465 North Halstead Street, Suite 140, Pasadena, CA 91107, USA}
	
	\affiliation{$*$Corresponding author: grudinin@jpl.nasa.gov}

\begin{abstract}
Kerr frequency combs with depressed harmonic at the optical pump frequency are theoretically explained and experimentally demonstrated. This
result is achieved in a MgF$_2$ photonic belt resonator having reduced density of modes in its spectrum and configured with the add-drop optical couplers. \copyright 2016 California Institute of Technology. Government sponsorship acknowledged.
\end{abstract}

\maketitle

Ring resonators characterized with cubic nonlinearity produce optical
frequency combs when pumped with continuous wave (cw) light of sufficient
power and certain frequency \cite{kippenberg11s,savchenkov16nano}.  These
frequency combs, dubbed as Kerr combs, may correspond to a train of short
optical pulses \cite{matsko11ol,herr14np,brash16s} and can be as broad as
an octave \cite{delhaye11prl,okawachi11ol}. In the majority of
experimental studies the resonator output contains the spectral line corresponding to the pump,
which is significantly stronger than all other lines of the mode-locked
frequency comb. The strong cw pump produces a high output cw background, which  makes it hard to observe various Kerr frequency comb regimes and the high contrast optical pulses emerging from the resonator. The frequency comb without the cw background is useful for radio frequency (RF) photonic filters and oscillators, fiber telecommunications, and spectrometers.

The large difference between the pump harmonic and the comb envelope, or
small contrast of the frequency comb, occurs in part due to saturation of
the resonant nonlinear process. Another reason is the geometrical
mismatch between the pump light and the mode of the cavity. While the
geometrical mismatch can be reduced, conventional spectrally broad Kerr
frequency combs still have low nonlinear conversion efficiency
\cite{bao14ol,yi15o}, so that most of the pump power is reflected from the resonator during comb generation. This happens because the laser frequency has to be tuned
away from the resonator mode to support generation of the mode-locked Kerr frequency comb.

The add-drop coupler configuration, consisting of an input coupler (add port) and an output coupler (drop port), allows reducing the pump spectral line to the comb envelope level by
the mode anti-crossing approach\cite{wang13oe}. Usually the pump line still exceeds the harmonics of the mode-locked Kerr comb retrieved from the drop port rather significantly \cite{liang15np,wang16arch}. It
is desirable to find a regime of mode-locked Kerr frequency comb
generation that does not suffer from the excessive cw background. Moreover, the theory describing frequency comb generation in microresonators predicts the existence of such a regime, which has not been observed so far. One of the main results of our study is that the theory is complete and the observation of the regime is possible if a proper ring resonator is made. We show both theoretically and experimentally that it is feasible to achieve comb generation regimes for which the intracavity comb envelope exceeds the
pump line, resulting in a smooth comb envelope at the drop port,
where the pump reflected from the resonator is not present.

Our paper is organized as follows. We first derive the analytical formula
showing the condition for the ideal intracavity comb envelope to exceed
the pump line. Then we validate the analytical result numerically and
present the Kerr frequency comb with a suppressed pump line
experimentally generated with a MgF$_2$ photonic belt resonator. We compare the experimental results with numerical simulations and provide an explanation of why pump harmonic suppression is not present if a large number of high order modes are supported by a resonator.

We consider a mode locked Kerr frequency comb as a train of $N$ identical
temporal dissipative Kerr solitons circulating in the cavity. The
envelope amplitude of the pulses can be found from the damped driven
Nonlinear Shr\"odinger equation also known as Lugiato-Lefever equation
(LLE). We introduce the dimensionless parameters for the detuning of the
pump line $\omega$ from the frequency of the corresponding resonator
mode $\omega_0$, $\zeta_0=(\omega_0-\omega)/\gamma$, and for the pump
power $f^2=(2g\eta P_\mathrm{in})/(\gamma^2\hbar\omega_0)$.
Here $P_\mathrm{in}$ is the input power,
$g=(\hbar\omega_{0}^{2}cn_{2})/(n^{2}V_{\mathrm{eff}})$ is the nonlinear
coupling constant, which is also the Kerr frequency shift per photon
\cite{matsko05pra}, $V_{eff}$ is the mode volume, $n$ and $n_2$ are linear and nonlinear
refraction indices, $c$ is the speed of light, $\hbar$ is the reduced Planck's
constant,
and $\eta=1-\gamma_0/\gamma$ is the coupling efficiency parameter with
$\gamma$ and $\gamma_0$ corresponding to total and intrinsic resonance
half-linewidths. The group velocity dispersion (GVD) is characterized by
$d_2=(\omega_{+1}+\omega_{-1}-2\omega_0)/(2\gamma)$, where $\omega_{\pm 1}$
are the frequencies of the modes neighboring with the pumped mode. We
approximate the solution of LLE as \cite{herr14np,matsko13oe,nozaki86}
\begin{align}
u(\phi)\simeq\left [c_0+ se^{i\chi} \sum\limits_{j=1}^N \sech\; b(\phi-\phi_j)
\right ]e^{i\psi_0},
\end{align}
where $\phi$ is the angular coordinate in the frame that co-propagates with the field inside the ring resonator, and $\phi_j$ are soliton positions. Real $c_0\simeq f/\zeta_0$ and $\psi_0$ stand
for cw constant background amplitude and phase. These values can be found from the algebraic stationary equation which is obtained from the LLE by setting all temporal partial
derivatives to zero. The solitonic parameters can be found with the Lagrangian perturbation
method \cite{herr14np,matsko13oe} as $s\simeq (2\zeta_0)^{1/2}$, $b\simeq(\zeta_0/d_2)^{1/2}$,
$\chi \simeq \pi-\arcsin [ (8\zeta_0)^{1/2}/(\pi f) ]$.

%
%\begin{figure}[htbp]
\begin{figure}
  \centering
   \includegraphics[width=0.8\columnwidth]{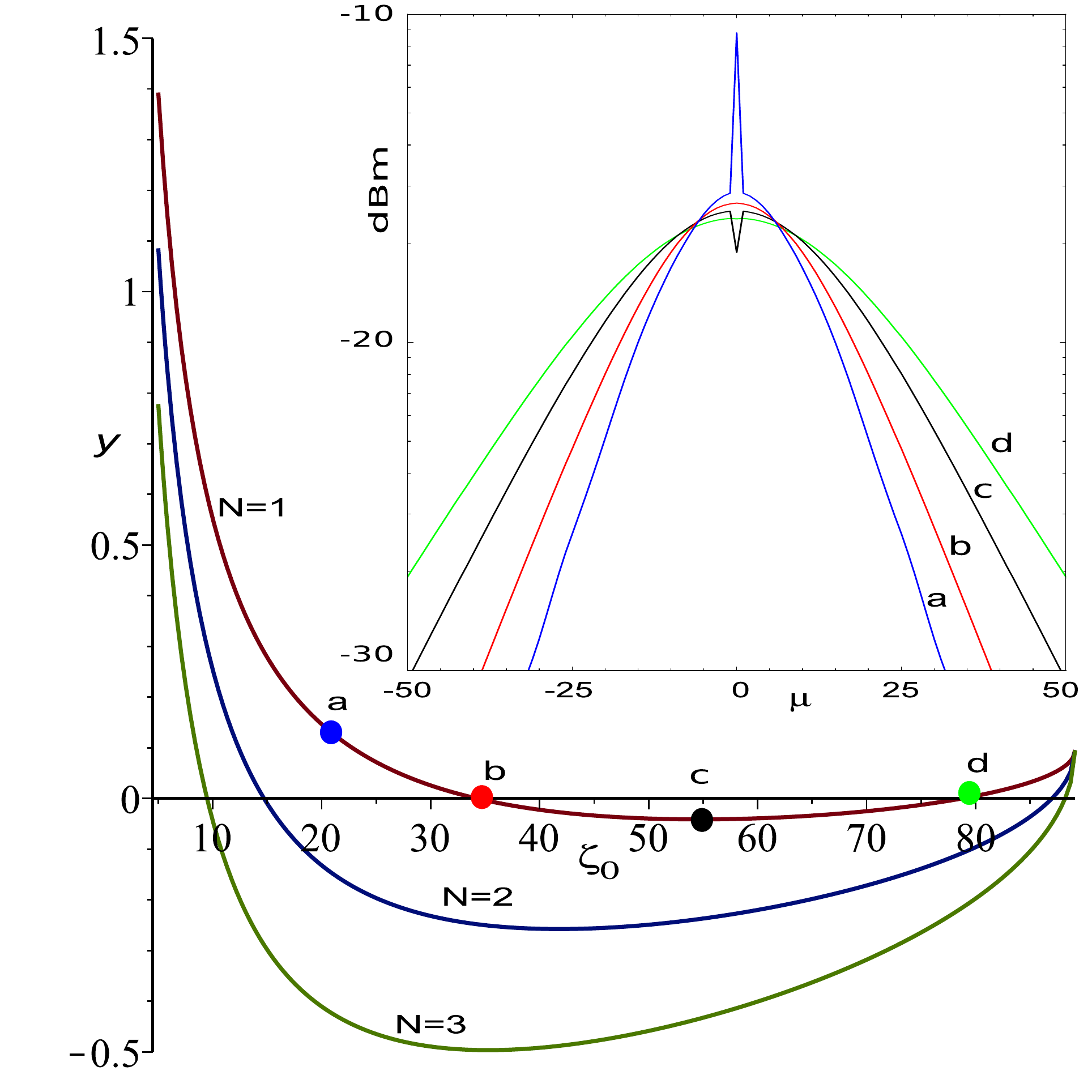}
\caption{ \small Numerically evaluated expression for the apparent carrier
suppression for $N=1,2,3$ and the single-soliton Kerr frequency combs
(inset) for four different detunings $\zeta_0$.
The pump harmonic is suppressed (b,d) at certain
detunings given by Eq.~(\ref{cond}).}
\label{fig:fig1}
\end{figure}

Considering that the width of the solitons is small ($2\pi b\ll 1$) we
find the cw component of the field:
\begin{align}
u_0={\cal
F}[u(\phi)](0)=&e^{i\psi_0}\int\limits_0^{2\pi}[c_0+s e^{i\chi}\sum\limits_{j=1}^N
\sech(b(\phi-
\phi_j))]\frac{d\phi}{2\pi}\nonumber\\ \simeq &(c_0+s_0e^{i\chi})e^{i\psi_
0}
=(c_0+\frac{Ns}{2b}e^{i\chi})e^{i\psi_0}
\end{align}
The condition for generation of the high contrast comb is thus $Ns/2b\geq c_0$ (the soliton's pump frequency Fourier component amplitude $s_0$ is larger than cw background amplitude $c_0$). It is interesting, however, that apparent suppression of the pump harmonic is also possible, when $|u_0|^2=s_0^2$, which leads to
\begin{align}
y=\frac{Ns}{b}\cos\chi+c_0=0
\end{align}
Note that in this case, though the amplitude of the pump line is the
same as that of soliton's pump harmonic, the phase is different, ($\pi-\chi$ instead of
$\chi$).
Substituting the approximated soliton parameters we arrive at the
condition of the apparent pump harmonic suppression:
\begin{align}
y=\frac{f}{\zeta_0}-N\sqrt{2 d_2}\sqrt{1-\frac{8\zeta_0}{\pi^2f^2}}=0
\label{cond}
\end{align}

An example of graphical solution of the above equation obtained by varying
the relative detuning $\zeta_0$  and using $f=8.5$, $d_2=0.05$ is shown
in Fig.\ref{fig:fig1}. This also includes the simulated single-soliton
Kerr comb spectra corresponding to several values of $\zeta_0$. Numerical
simulations were performed using mode decomposition approach \cite{matsko05pra,chembo10pra,herr14prl}. Exploring 
multiple parameters of the system we found that the carrier suppression is 
always possible if the pump power is high enough. Similar result was predicted 
theoretically for the case of a modulation instability laser \cite{coen98oc}.
\begin{figure}[!ht]
  \centering
 \includegraphics[width=\columnwidth]{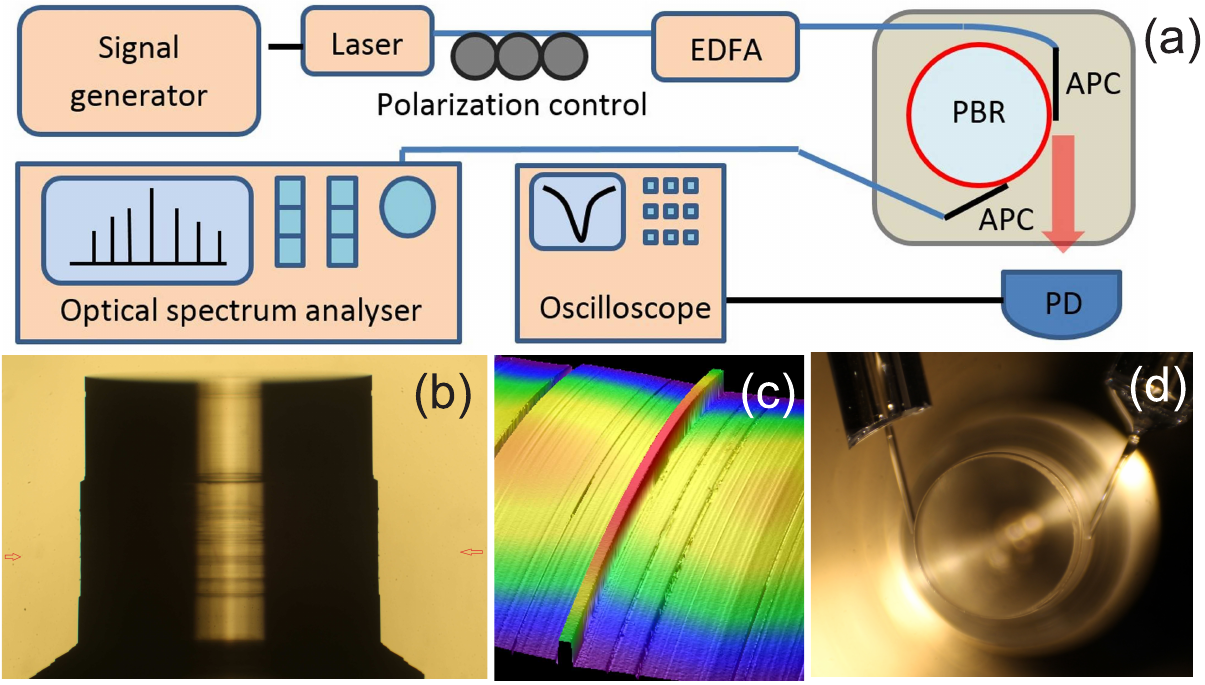}
\caption{ \small (a) Schematics of the experimental setup. Light emitted by a
tunable fiber laser is amplified with an erbium-doped fiber amplifier
(EDFA) and coupled into a photonic belt resonator (PBR) using a mode-matched angle-polished fiber coupler (APC). Light exiting the resonator
through the first APC is detected by a photodiode (PD). Light from
another APC (drop port) is measured with an optical spectrum analyzer
(Yokogawa AQ6319). (b) Picture of the cylinder that holds the PBR
structures. Several belt resonators are made on the side surface of
the crystalline cylinder by single point diamond turning. (c) Profile of the
PBR taken with an optical interferometric profilometer Veeco Wyko NT9300. (d)
Coupling of two APC (add and drop ports) with the PBR.
} \label{fig:fig2}
\end{figure}

\begin{figure}
\centering %\sidecaption
\includegraphics[width=\columnwidth]{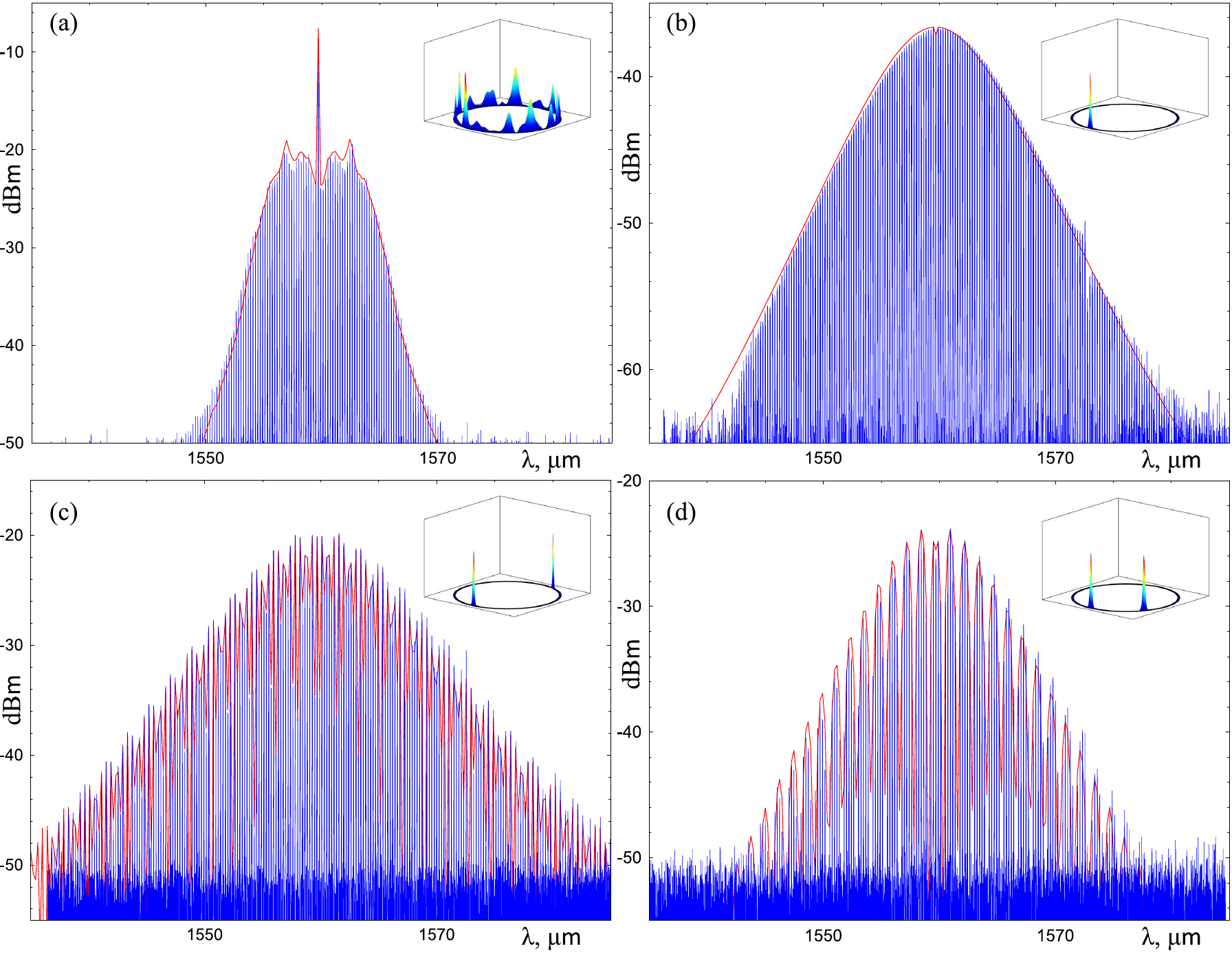}
\caption{\small Numerically simulated optical spectra (shown in red) compared with the
experimental results (shown in blue).
(a) Incoherent spectrum. To obtain the smooth spectrum envelope we
averaged the time-dependent chaotic spectrum
over several cavity ring-down times. This procedure resembles the
operation of a conventional optical spectrum analyzer \cite{matsko13ol}. (b-d) Kerr comb soliton
spectra. The simulations are performed for $f=8.5$ and $d_2=0.05$. This
corresponds to the parameters of our MgF$_2$ comb generator:
$P=19$ mW, $Q=\omega_0/2\gamma=0.47\times10^9$, $\eta=1/2$, $n=1.37$,
$\lambda=1.561$ $\mu$m, $n_2=0.9\times10^{-20}$(m/V)$^2$,
$V_\mathrm{eff}=2.8\times 10^{-13}$m$^3$. }
\label{fig:fig3}
\end{figure}

To validate the theoretical prediction we fabricated the crystalline resonator with only one mode per free spectral range (FSR) \cite{grudinin15o}, and coupled this resonator in add-drop configuration \cite{wang13oe,liang15np,wang16arch}
(Fig.~\ref{fig:fig2}). We pumped this resonator with 1561~nm light and observed generation of Kerr frequency combs with the suppressed pump harmonic (Fig.~\ref{fig:fig3}).

The experimental setup is shown in Fig.~(\ref{fig:fig2}a). Three photonic
belt resonators (PBRs) are fabricated on a MgF$_2$ polished cylinder
with 2675 micrometers in diameter Fig.~(\ref{fig:fig2}b). These
resonators have typical cross sections on the order of $10\times 10$
micrometers and support dispersion
engineering  Fig.~(\ref{fig:fig2}c) \cite{grudinin15o}. Angle polished
SMF-28 fiber couplers were used for add and drop ports. The coupling efficiency is measured as the maximum achievable dip of transmission when the distance between the coupler and the PBR is varied. This minimum transmission corresponds to critical coupling, where the resonator’s coupling loss equals the intrinsic loss. The critical coupling can be less than 100\% if resonator’s output beam is not matched well to the coupler transmission. Here, all measurement were done under critical coupling setting and the transmission dip was over 60\%. The drop port coupling was only enabled during comb measurements. The additional small coupler loss results in the shift towards the under-coupled regime \cite{ideality}. We
measured the Q factor of a specific PBR at 470 million (critically
coupled) by recording its resonance curve with a Koheras 1561~nm laser
and using phase modulation sidebands for frequency calibration. While single-mode by design, this resonator still supports a couple of higher order modes with poor coupling efficiency and low Q. These modes are well separated from the fundamental
mode frequency in the vicinity of the pump wavelength.
%
%\begin{figure*}[!ht]
%\end{figure*}
%
We excited the frequency comb in this resonator by pumping it with
19~mW of light as measured at the output of the input angle-polished fiber coupler (``add'' port).
We observed the step-like multistability on the resonance curve, which is
typical for soliton combs \cite{herr14np,herr14prl,brash16s}, by tracking the power of the light reflected from the resonator (``add'' port) as the laser frequency was scanned through the resonator mode. 

The laser frequency was first tuned to the blue slope of the resonance
and we observed generation of the first pair of sidebands at 11 FSRs of the resonator (FSR$=25.78$~GHz). The laser was
then tuned closer to the mode center and a comb  typical of a non-coherent state was observed (Fig.~\ref{fig:fig3}a) \cite{matsko13ol}. We then used the frequency jump technique to excite the comb states that are
typical of single and multiple soliton states (Fig.~\ref{fig:fig3}b-d).
In this technique, the laser frequency is tuned abruptly to the expected
solitonic state within the nonlinear resonance curve. As predicted by our calculations, we found that the pump harmonic was suppressed in all of these states and merged with the overall comb envelope. No tuning of experimental parameters was required. In our case the soliton states were not
very stable due to temperature drift of the resonator after the frequency
jumps. The slight asymmetry of the recorded spectra in addition to higher
order dispersion may be also attributed to the drift of the detuning
between the pump light and the corresponding resonator mode during the
time needed to acquire such single spectrum (10-30~s). Further work is
needed to stabilize the frequency comb completely and can be achieved by
controlling the laser parameters after the frequency jumps.

There is a very good agreement between the experimental data and the results of our numerical simulations with actual experimental parameters (Fig.~\ref{fig:fig3}). The evaluation  started at a chosen initial
detuning with seeded analytical soliton train and propagated in time with
slow pump frequency tuning until the boundary of soliton existence was
reached. The number of calculated modes was selected to be $255$. The
case of anomalous group velocity dispersion corresponding to the
experimental conditions was considered. The pump frequency detuning was adjusted until simulated comb width matched the experimental one. It was assumed that the
intracavity field is sampled using the drop coupler, so the power of all
harmonics of the frequency comb outside of the resonator is proportional
to the intracavity power.

We found that some experimental spectra correspond to two optical pulses
confined in the mode. In order to simulate the two-soliton spectra we
determined the angular distance between the solitons $\phi_2-\phi_1$ by
means of Fourier transform of the experimental spectra. Such
transformation produces the autocorrelation function for the waveform
\cite{brash16s}. This angular distance was then used to generate the
analytical expression for the soliton train, which was then used as an
initial condition in the code. We found that the resulting frequency comb
was not stable for smaller detunings (corresponding to longer solitons)
when the experimentally determined distance was too small
(Fig.~\ref{fig:fig3}d). The solitons were interacting, moving with
respect to each other, until a stationary two-soliton cluster was formed.
This observation explains the experimental difficulties with the
observation of these particular frequency comb states as well as the
small dissimilarity of the wings of the comb envelope observed
experimentally and simulated numerically (Fig.~\ref{fig:fig3}d).  

Let us now discuss the reasons why the pump harmonic is not always
completely suppressed if one uses the add-drop port configuration and
generates a broad mode-locked Kerr frequency comb. In this regime the
conversion efficiency of the nonlinear process is saturated and the
intracavity field at the pump frequency does not increase
significantly when the pump power exceeds the oscillation threshold. Most
of the pump light is reflected from the resonator. This reflection does
not impact the light leaving the drop port in the add-drop coupler
configuration if the resonator has a single mode family. Therefore, the
carrier is expected to be suppressed in a single mode resonator.
However, the insufficient pump power is one of the reasons why it does
not always happen. Indeed, let us note that the condition of high
contrast solitons $\sqrt{d_2}\zeta_0/2f>1$ (which follows from (2) and $s/2b\geq c_0$) may be cast as $\sqrt{d_2}
f\gtrsim 1$ if we take the soliton existence condition into account:
$\zeta_0\leq \pi^2 f^2/8$ \cite{herr14np, nozaki86}. It means that the
pump power needed for the suppression of the carrier harmonic is
inversely proportional to $Q^3$ of the mode (since, by definition, 
$d_2\sim Q$ and $f^2\sim P_{in} Q^2$). This value is too large to
be achieved in the majority of experiments involving low-Q resonators.
Barring a few weakly coupled higher order low-Q modes, the resonator used
in this work has the highest Q-factor of any single mode family
resonators studied to date.

Dense spectrum of resonator modes is another reason why the pump harmonic suppression has
not been experimentally observed in typical multimode resonators so far.
There are multiple channels for the pump light to go from the add to the
drop port in the case of a multimode resonator. The leakage through the
higher order modes with nearly the same frequency as the pumped mode is
suppressed by mode matching of the couplers. However, this
suppression usually does not exceed 20~dB. Some pump light enters the
high order mode families and leaves the resonator without participating
in the nonlinear frequency conversion. As a result, the output power at
the carrier frequency increases, in contrast with the power leaving a
single mode family resonator, where such an interaction is not supported.

\begin{figure}
  \centering
\includegraphics[width=0.8\columnwidth]{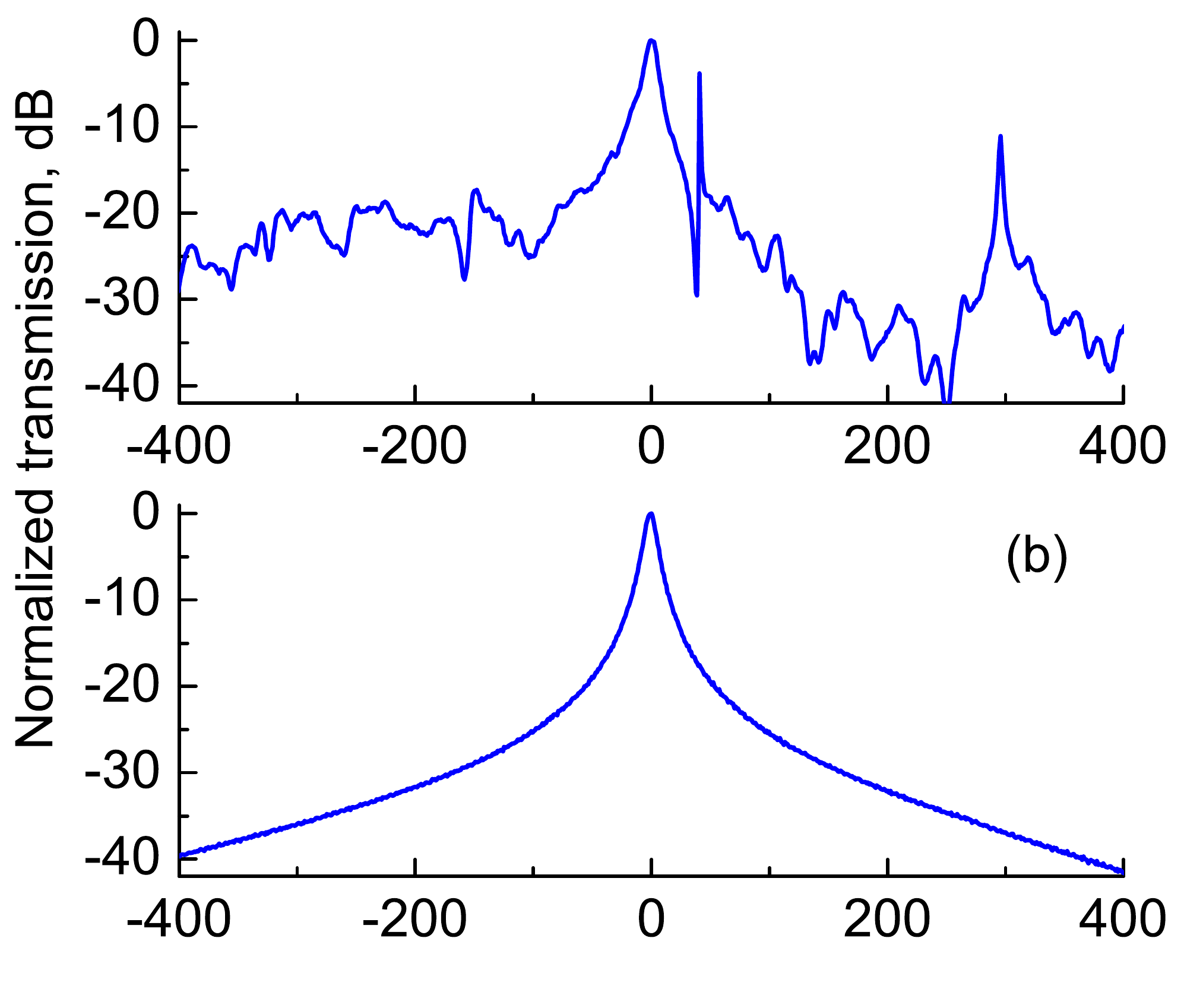}
\caption{ \small Linear optical spectra of a sharp edge multimode, (a), and a
single mode photonic belt, (b), resonators. The spectra are taken in the
add-drop port configuration.
} \label{fig:fig4}
\end{figure}

To illustrate this process we have measured the modal spectra of two
different resonators in the linear regime. The spectrum of an 18~GHz FSR wedge-shaped resonator is shown in (Fig.~\ref{fig:fig4}a). The resonator supports a large number of high order modes that create a -20~dB background in the spectrum. Moreover,
there is a mode that nearly overlaps with the pump mode, resulting in interference and
sharp dispersion features (Fano resonances
\cite{maleki04ol,li11apl}). When Kerr frequency comb is generated with such
a resonator, up to ten percent of the pump light reaches the drop port through the adjacent modes.
Similarly, this explains the large pump line on top of the mode locked Kerr comb generated with multimode resonators in some studies.
It is worth noting that the leakage light from the higher order modes can result in multi-path interference noises in the detected beatnote signals.
On the other hand, there is only one channel for the pump light to reach the drop port (Fig.~\ref{fig:fig4}b) in the single mode belt resonator \cite{grudinin15o, winkler16oe}. Observation of the intra-cavity field becomes
possible, as the result.

In conclusion, we have studied  generation of Kerr frequency combs in
a nonlinear ring resonator pumped with cw coherent light under conditions
optimized for increased comb contrast. We have shown that it is
possible to observe Kerr frequency combs with suppressed pump line
when the density of the optical modes in the resonator is reduced.
Any increase of the mode density results in a decrease of the Kerr comb
contrast due to leakage of the pump light through  high-order modes
not involved in the nonlinear process. The observation is important for
multiple practical applications of the microresonator-based frequency
combs suffering from the presence of the large cw background produced by the pump harmonic which masks
optical pulses leaving the resonator and increases the noise of the comb
oscillators.

%\section{Funding Information}
The research was carried out at the Jet Propulsion Laboratory, California Institute of Technology, under a contract with the National Aeronautics and Space Administration. We thank Risaku Toda of JPL's MDL for the optical
profilometer images. M.L.G acknowledges support from the Ministry of Education and Science of
the Russian Federation project RFMEFI58516X0005. V.H. acknowledges support from the Direction Générale de l’Armement (DGA). A.B.M. and L.M. acknowledge support from Defense Sciences Office of Defense Advanced Research Projects Agency under contract D16PC00158.


\begin{thebibliography}{99}


\bibitem{kippenberg11s}  T. J. Kippenberg, R. Holzwarth, and S. A. Diddams, ``Microresonator-based optical frequency combs,'' Science {\bf 332}, 555-559 (2011).

\bibitem{savchenkov16nano} A. A. Savchenkov, A. B. Matsko, and L. Maleki, ``On frequency combs in monolithic resonators,'' Nanophotonics {\bf 5}, 363-391 (2016).

\bibitem{matsko11ol} A. B. Matsko, A. A. Savchenkov, W. Liang, V. S. Ilchenko, D. Seidel, and L. Maleki,  ``Mode-locked Kerr frequency combs,''	Opt. Lett. {\bf 36}, 37 (2011).

\bibitem{herr14np} T. Herr, V. Brasch, J. D. Jost, C. Y. Wang, N. M. Kondratiev, M. L. Gorodetsky, and T. J. Kippenberg, ``Temporal solitons in optical microresonators,'' Nature Photon. {\bf 8}, 145 (2014).

\bibitem{brash16s} V. Brasch, M. Geiselmann, T. Herr, G. Lihachev, M . H. P. Pfeiffer, M. L. Gorodetsky, and T. J. Kippenberg, ``Photonic chip-based optical frequency comb using soliton Cherenkov radiation,'' Science {\bf 351}, 357 (2016).

\bibitem{delhaye11prl} P. Del'Haye, T. Herr, E. Gavartin, M. Gorodetsky, R. Holzwarth, and T. Kippenberg, ``Octave spanning tunable frequency comb from a microresonator,'' Phys. Rev. Lett. {\bf 107}, 063901 (2011).

\bibitem{okawachi11ol} Y. Okawachi, K. Saha, J. S. Levy, Y. H. Wen, M. Lipson, and A. L. Gaeta, ``Octave-spanning frequency comb generation in a silicon nitride chip,'' Opt. Lett. {\bf 36}, 3398 (2011).

\bibitem{yi15o} X. Yi, Q.-F. Yang, K. Y. Yang, M.-G. Suh, and K. Vahala, ``Soliton frequency comb at microwave rates in a high-Q silica microresonator,'' Optica {\bf 2}, 1078 (2015).

\bibitem{bao14ol} C. Bao, L. Zhang, A. Matsko, Y. Yan, Z. Zhao, G. Xie, A. M. Agarwal, L. C. Kimerling, J. Michel, L. Maleki, and A. E. Willner, ``Nonlinear conversion efficiency in Kerr frequency comb generation,'' Opt. Lett. {\bf 39}, 6126 (2014).


\bibitem{wang13oe} P.-H. Wang, Y. Xuan, L. Fan, L. T. Varghese, J. Wang, Y. Liu, X. Xue, D. E. Leaird, M. Qi, and A. M. Weiner, ``Drop-port study of microresonator frequency combs: power transfer, spectra and time-domain characterization,'' Opt. Express {\bf 21}, 22441 (2013).

\bibitem{liang15np} W. Liang, D. Eliyahu, V. S. Ilchenko, A. A. Savchenkov,	A. B. Matsko, D. Seidel, and L. Maleki, ``High spectral purity Kerr frequency comb radio frequency photonic oscillator,'' Nature Communications  {\bf 6}, 7957 (2015).

\bibitem{wang16arch} P. H. Wang, J. A. Jaramillo-Villegas, Y. Xuan, X. Xue, C. Bao, D. E. Leaird, M. Qi, A. M. Weiner,  ``Intracavity characterization of micro-comb generation in the single-soliton regime,''	$arXiv:1603.03154$ (2016).

\bibitem{matsko05pra} A. B. Matsko, A. A. Savchenkov, D. Strekalov, V. S. Ilchenko, and L. Maleki, ``Optical hyperparametric oscillations in a whispering-gallery-mode resonator: Threshold and phase diffusion,'' Phys. Rev. A {\bf 71}, 033804 (2005).

\bibitem{matsko13oe} A. B. Matsko and L. Maleki, ``On timing jitter of mode locked Kerr frequency combs,'' Opt. Express 21, 28862 (2013).	

\bibitem{nozaki86} K. Nozaki, N. Bekki, ``Low-dimensional chaos in a driven damped nonlinear shrodinger-equation,'' Physica D {\bf 21}, 381 (1986).

\bibitem{chembo10pra} Y. K. Chembo and N. Yu, ``Modal expansion approach to optical-frequency-comb generation with monolithic whispering-gallery-mode resonators,'' Phys. Rev. A {\bf 82}, 033801 (2010)..

\bibitem{herr14prl} T.Herr, V. Brasch, J. D. Jost, I. Mirgorodskiy, G. Lihachev, M. L. Gorodetsky, and T. J. Kippenberg, ``Mode spectrum and temporal soliton formation in optical microresonators,'' Phys. Rev. Lett. {\bf 113}, 123901 (2014).

\bibitem{coen98oc} S. Coen, M. Haelterman, ``Impedance-matched modulational instability laser for background-free pulse train generation in the THz range,'' Opt. Commun. {\bf 146}, 339 (1998).	


\bibitem{grudinin15o} I. S. Grudinin and N. Yu, ``Dispersion engineering of crystalline resonators via microstructuring,'' Optica {\bf 2}, 221	(2015).

\bibitem{ideality} S. M. Spillane, T. J. Kippenberg, and K. J. Vahala, ``Ideality in a fiber-taper-coupled microresonator system for application to cavity quantum electrodynamics,'' Phys. Rev. Lett. {\bf 91} 043902 (2003).


\bibitem{matsko13ol} A. B. Matsko, W. Liang, A. A. Savchenkov, and L. Maleki, ``Chaotic dynamics of frequency combs generated with continuously pumped nonlinear microresonators,'' Opt. Lett. {\bf 38}, 525 (2013).

\bibitem{maleki04ol} L. Maleki, A. B. Matsko, A. A. Savchenkov, and V. S. Ilchenko, ``Tunable delay line with interacting whispering-gallery-mode resonators,'' Opt. Lett. {\bf 29}, 626 (2004).

\bibitem{li11apl} B. B. Li, Y. F. Xiao, C. L. Zou, Y. C. Liu, X. F. Jiang, Y. L. Chen, Y. Li, and Q. Gong, ``Experimental observation of Fano resonance in a single whispering-gallery microresonator,'' Appl. Phys. Lett. {\bf 98}, 021116 (2011).

\bibitem{winkler16oe} J. M. Winkler, I. S. Grudinin and N. Yu, ``On the properties of single-mode optical resonators,'' Opt. Express {\bf 24}, 13231 (2016).

\end{thebibliography}
\end{document}